\documentclass{article}
\usepackage{amsmath}

\usepackage{amssymb, amscd, amsfonts, csquotes, graphicx, caption, fullpage, float, epstopdf, color, subcaption , setspace, tensor, url, verbatim, lipsum, blkarray}
\usepackage[final]{hyperref}
\usepackage{cleveref}
\usepackage[table,xcdraw]{xcolor} % for \colot[HTML] 
\usepackage[linesnumbered,ruled,noend]{algorithm2e} %for algorithms
\usepackage{multirow}
\usepackage{amsmath,amsthm}
\usepackage{mdframed}
\usepackage[flushleft]{threeparttable} % http://ctan.org/pkg/threeparttable
\usepackage{longtable}
\usepackage{booktabs}
\usepackage{xargs}                      % Use more than one optional parameter in a new commands
\usepackage{pgf}
\usepackage{tikz}
\usetikzlibrary{arrows,automata,positioning}
\usepackage{listings}
\usepackage{lstautogobble} % Fix relative indenting
\usepackage{color} % Code coloring
\usepackage{zi4} % Nice font
\definecolor{bluekeywords}{rgb}{0.13, 0.13, 1}
\definecolor{greencomments}{rgb}{0, 0.5, 0}
\definecolor{redstrings}{rgb}{0.9, 0, 0}
\definecolor{graynumbers}{rgb}{0.5, 0.5, 0.5}
\definecolor{mycolor}{RGB}{0, 124, 255}
\definecolor{mycolornavy}{RGB}{0, 35, 102}

\global\mdfdefinestyle{framing}{%
linecolor=black,linewidth=2pt,%
leftmargin=1cm,rightmargin=1cm
}

\title{FCMpy: {\ttfamily Python} Module for Constructing and Analyzing Fuzzy Cognitive Maps}

\author{Samvel Mkhitaryan\\\textit{Maastricht University}\\
    Philippe J. Giabbanelli $\And$ Maciej K. Wozniak\\\textit{Miami University}\\
   Gonzalo Nápoles\\\textit{Tilburg University}\\
   Nanne K. de Vries $\And$ Rik Crutzen\\\textit{Maastricht University}}
   
\begin{document}
\maketitle
\begin{abstract}
    FCMpy is an open source package in \textit{Python} for building and analyzing Fuzzy Cognitive Maps.More specifically, the package allows 1) deriving fuzzy causal weights from qualitative data, 2) simulating the system behavior, 3) applying machine learning algorithms (e.g., Nonlinear Hebbian Learning, Active Hebbian Learning, Genetic Algorithms and Deterministic Learning) to adjust the FCM causal weight matrix and to solve classification problems, and 4) implementing scenario analysis by simulating hypothetical interventions (i.e., analyzing what-if scenarios).
\end{abstract}

%% -- Introduction -------------------------------------------------------------

\section[Introduction]{Introduction} \label{sec:intro}
Fuzzy Cognitive Maps (FCM) were introduced by B. Kosko as an extension to the traditional cognitive maps and are used to model and analyze complex systems \cite{kosko:1986, Axelrod:2015}. FCMs are applied in a variety of fields such as engineering \cite{Stylios:2004}, health sciences and medicine \cite{Papakostas:2011, Salmeron:2012, Giabbanelli:2012}, environmental sciences \cite{Kok:2009, Papageorgiou:2012}, and political analysis \cite{Andreou:2005, Giabbanelli:2014}.

An FCM represents a system as a directed signed graph where components are represented as nodes and the causal relationships between these components are represented by weighted directed edges. The dynamics of the system are examined by simulating its behavior over discrete simulation steps. In general, FCMs can be constructed based on the inputs of domain experts (i.e., expert based FCMs), data collected about the system (e.g., data driven approaches) or the combination of the two (i.e., hybrid approaches) \cite{Mkhitaryan:2020}. 

The available solutions for constructing and analyzing FCMs come in the form of dedicated software solutions (e.g., \emph{Mental Modeler}, \emph{FCM Designer}), open source libraries (e.g., \textit{fcm} package available in \textit{R}, \textit{pfcm} available in \textit{Python}) and open source scripts \cite{Firmansyah:2019, Napoles:2018}. However, the available open source solutions provide only partial coverage of the useful tools for building and analyzing FCMs, lack generality for handling different use cases, or require modifying the source code to incorporate specific features \cite{Napoles:2018}. For example, the \textit{fcm} and \textit{pfcm} provide utilities for simulating FCMs but not for constructing them based on qualitative (e.g., by applying fuzzy logic) or quantitative inputs. To our knowledge, none of the available open source solutions available (e.g., \textit{R} and \textit{Python}) implements machine learning algorithms for FCMs (e.g., NHL, AHL, RCGA). Although several software packages have successfully implemented such algorithms (e.g., FCM Expert, FCM Wizard), their reliance on a graphical user interface prevents their integration in a data science workflow articulated around a language such as Python \cite{napoles2018fcm}.

The dedicated modules in our proposed \textit{FCMpy} package provide utilities for 1) constructing FCMs based on qualitative input data (by applying fuzzy logic), 2) simulating the system behavior, 3) implementing machine learning algorithms (e.g.,  Nonlinear Hebbian Learning, Active Hebbian Learning, Genetic Algorithms and Deterministic Learning) to optimize the FCM causal weight matrix and model classification problems, and 4) implementing scenario analysis by simulating hypothetical interventions (i.e., analyzing what-if scenarios).

%% -- Manuscript ---------------------------------------------------------------

%% - In principle "as usual" again.
%% - When using equations (e.g., {equation}, {eqnarray}, {align}, etc.
%%   avoid empty lines before and after the equation (which would signal a new
%%   paragraph.
%% - When describing longer chunks of code that are _not_ meant for execution
%%   (e.g., a function synopsis or list of arguments), the environment {Code}
%%   is recommended. Alternatively, a plain {verbatim} can also be used.
%%   (For executed code see the next section.)

\section{Constructing Expert-Based FCMs} \label{sec:ExpertFCM}
Expert-based FCMs are often constructed based on data collected from the domain experts (e.g., by the means of surveys) where the domain experts first identify the factors relevant to the problem domain and then express the causal relationships between these factors with linguistic terms (e.g., very high, high, low). Fuzzy logic is subsequently applied to convert linguistic ratings into numerical weights (i.e., crisp values). The conversion of linguistic ratings to numerical weights includes the following four steps \cite{Mkhitaryan:2020, mago2012supporting, mago2013analyzing}: 1) define fuzzy membership functions for the linguistic terms, 2) apply fuzzy implication rule onto the fuzzy membership functions based on the expert ratings, 3) combine the membership functions resulting from the second step with an aggregation operation, and 4) defuzzify the aggregated membership functions. In this section, we first describe methods for reading data from different file formats and then describe the methods for constructing expert-based FCMs based on qualitative data.

\subsection{Data handling}
The available open source solutions for expert-based FCMs do not provide utilities for working with different file types thus limiting their usability. Data on FCMs include the edges (represented as pairs of source/target) and the associated linguistic ratings of the survey participants. The \emph{ExpertFcm} class provides a {\ttfamily $read\_data()$} method for reading data from .csv, .xlsx, and .json files (see the code snippet below). The corresponding files should satisfy certain requirements that are described in detail in the PyPI documentation. The {\ttfamily $read\_data()$} requires the file path as an argument. The additional arguments that depend on the file extension (e.g., csv, json, xlsx) should be specified as keyword arguments. For the .xlsx and .json files, when the optional {\ttfamily $check\_consistency$} argument is set to {\ttfamily True} then the algorithm checks whether the experts rated the causal impact of the edges (source-target pairs) consistently in terms of the valence of the causal impact (positive or negative causality). If such inconsistencies are identified, the method outputs a separate .xlsx file that documents such inconsistencies.

\begin{lstlisting}[language=Python]
>>> from fcmpy import ExpertFcm

>>> fcm = ExpertFcm()

>>> data = fcm.read_data(file_path, 
                      sep_concept='->', csv_sep=';')
\end{lstlisting}

The {\ttfamily $read\_data()$} method returns an ordered dictionary where the keys are the experts' IDs (or the names of the excel sheets in the case of an excel file or the row index in case of a csv file) and the values are \textit{pandas} dataframes with the expert inputs (see the snippet of the code output below).

\begin{lstlisting}[language=Python]
OrderedDict(
[('Expert0',
    -vh  -h  -m  -l  -vl  na  +vl  +l  +m  +h  +vh From  To
0    1   0   0   0    0   0    0   0   0   0    0   C1  C2
1    0   0   0   0    0   0    0   0   0   1    0   C2  C1
2    0   0   0   0    0   0    0   0   0   1    0   C3  C1
3    0   0   0   0    0   0    1   0   0   0    0   C3  C4),
....
('Expert5',
    -vh  -h  -m  -l  -vl  na  +vl  +l  +m  +h  +vh From  To  no causality
0    0   0   1   0    0   0    0   0   0   0    0   C1  C2           0.0
1    0   0   0   0    0   0    0   0   1   0    0   C2  C1           0.0
2    0   0   0   0    0   0    0   0   0   0    0   C3  C1           1.0
3    0   0   0   0    0   0    0   0   0   0    0   C3  C4           1.0)]
)
\end{lstlisting}

It is often useful to check the extent to which the participants agree on their opinions with respect to the causal relationships between the edges. This is often done by calculating the information entropy \cite{Giabbanelli:2012, Firmansyah:2019} expressed as:

\begin{equation} \label{eq:entropy}
R=-\sum_{i=1}^{n}p_i log_2(p_i)
\end{equation}
where $p\_i$ is the proportion of the answers (per linguistic term) about the causal relationship. The entropy scores can be calculated with the {\ttfamily entropy()} method (see the code snippet below).

\begin{lstlisting}[language=Python]
>>> entropy = fcm.entropy(data)
\end{lstlisting}

\begin{lstlisting}[language=Python]
Entropy
From	  To	
C1	    C2	1.459148
C2	    C1	1.459148
C3	    C1	1.251629
C4	1.459148
\end{lstlisting}

\subsection{Four steps for obtaining causal weights}

To convert the qualitative ratings of the domain experts to numerical weights via fuzzy logic, we must 1) define the fuzzy membership functions, 2) apply a fuzzy implication rule, 3) combine the membership functions, and 4) defuzzify the aggregated membership functions to derive the numerical causal weights.

\subsubsection{Step 1: Define fuzzy membership functions}
Fuzzy membership functions are used to map the linguistic terms to a specified numerical interval (i.e., universe of discourse). In FCMs, the universe of discourse is specified in the range of [-1, 1] where the negative causality is possible or [0, 1] if otherwise. The universe of discourse can be specified with the {\ttfamily universe()} setter (see the code snippet below).

\begin{lstlisting}[language=Python]
>>> import numpy as np

>>> fcm.universe = np.arange(-1, 1.001, .001)
\end{lstlisting}

To generate the fuzzy membership functions we need to decide on the geometric shape that would best represent the linguistic terms. In many applications, a triangular membership function is used \cite{frias2017fuzzy}. The triangular membership function specifies the lower and the upper bounds of the triangle (i.e., where the meaning of the given linguistic term is represented the least) and the center of the triangle (i.e., where the meaning of the given linguistic term is fully expressed). 

The {\ttfamily $linguistic\_terms()$} method sets the linguistic terms and the associated parameters for the triangular membership function (see the code snippet below).

\begin{lstlisting}[language=Python]
>>> fcm.linguistic_terms = {
                        '-VH': [-1, -1, -0.75],
                        '-H': [-1, -0.75, -0.50],
                        '-M': [-0.75, -0.5, -0.25], 
                        '-L': [-0.5, -0.25, 0],
                        '-VL': [-0.25, 0, 0],
                        'No Causality': [-0.001, 0, 0.001],
                        '+VL': [0, 0, 0.25],
                        '+L': [0, 0.25, 0.50],
                        '+M': [0.25, 0.5, 0.75],
                        '+H': [0.5, 0.75, 1],
                        '+VH': [0.75, 1, 1]
                        }
\end{lstlisting}

\begin{figure}[H]
\centering
\includegraphics[width = \textwidth]{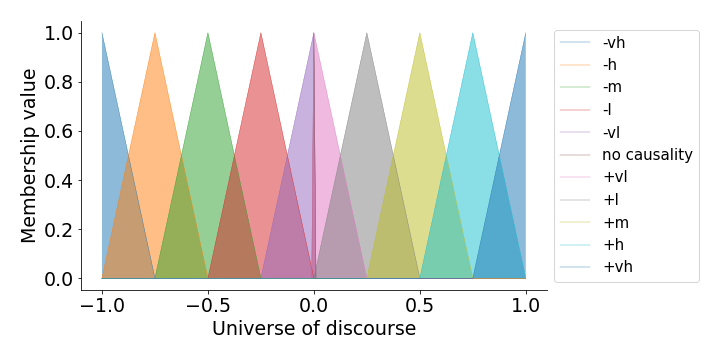}
\caption{\label{fig:trimf} Triangular membership functions}
\end{figure}

The keys in the above dictionary represent the linguistic terms and the values are lists that contain the parameters for the triangular membership function (i.e., the lower bound, the center and the upper bound) (see Figure \ref{fig:trimf}). After specifying the universe of discourse and the linguistic terms with their respective parameters one can use use the {\ttfamily automf()} method to generate the membership functions (see the code snippet below).

\begin{lstlisting}[language=Python]
>>> fcm.fuzzy_membership = fcm.automf(method='trimf')
\end{lstlisting}

In addition to the triangular membership functions, the {\ttfamily automf()} method also implements gaussian membership functions (`gaussmf') and trapezoidal membership functions ('trapmf') (based on \textit{sci-kit fuzzy} module in python).

\subsubsection{Step 2: Apply the fuzzy implication rule}

To determine the level of activation of the linguistic terms for a given pair of concepts, one must first identify the level of endorsement of the given terms by the participants. This is done by calculating the proportion of the answers to each linguistic term for a given edge. Consider a case where 50$\%$ of the participants (e.g., domain experts) rated the causal impact of an antecedent on the consequent as Positive High, 33$\%$ rated it as Positive Very High and the 16$\%$ rated it as Positive Medium. Subsequently, a fuzzy implication rule is used to "activate" the corresponding membership functions. Two such rules are often used, namely Mamdani's minimum and Larsen's product implication rule \cite{Nandi:2012, Stach:2005}. 

The Mamdani minimum fuzzy implication rule is expressed as:

\begin{equation} \label{eq:min}
\mu_{R}(x,y)= min \left \lfloor \mu_{A}(x), \mu_{B}(y) \right \rfloor
\end{equation}

where $\mu_{A}(x)$ and $\mu_{B}(y)$ denote the membership value x to the linguistic term A and the membership value y to the linguistic term B respectively.

\begin{figure}[H]
  \centering
  \subfloat[Mamdani]{\includegraphics[width=0.5\textwidth]{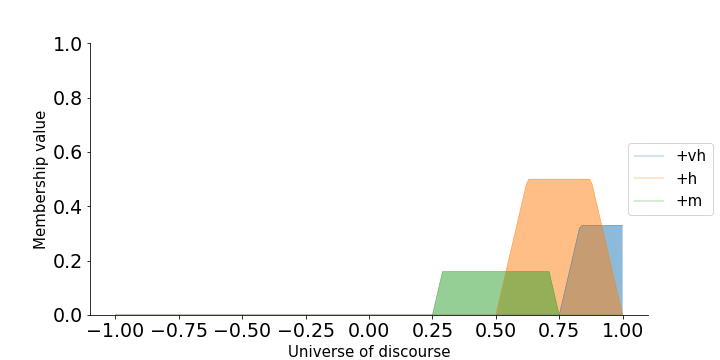}\label{fig:f1}}
  \hfill
  \subfloat[Larsen]{\includegraphics[width=0.5\textwidth]{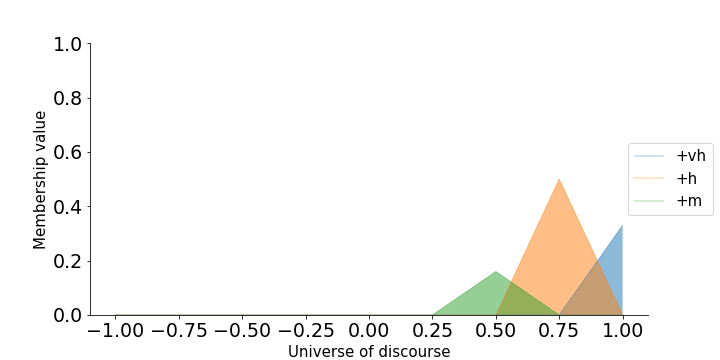}\label{fig:f2}}
  \caption{Fuzzy implication rules}
\end{figure}

The Mamdani rule cuts the membership function at the level of endorsement (see Figure \ref{fig:f1}). In contrast, Larsen's implication rule re-scales the membership function based on the level of endorsement (see Figure \ref{fig:f2}) and is expressed as: 

\begin{equation} \label{eq:prod}
\mu_{R}(x,y)= \mu_{A}(x)\cdot \mu_{B}(y)
\end{equation}

We can use {\ttfamily $fuzzy\_implication()$} method to apply the selected implication method (see the available methods in Table~\ref{tab:implication} and the code snippet below).

\begin{table}[t!]
\centering
\begin{tabular}{lllp{7.4cm}}
\hline
Argument         & Option & Description \\ \hline
{\ttfamily method}        & `Mamdani'  & Mamdani's fuzzy implication rule \\
                & `Larsen'  & Larsen's fuzzy implication rule \\\hline
\end{tabular}
\caption{\label{tab:implication} Fuzzy implication rules}
\end{table}

\begin{lstlisting}[language=Python]
>>> mfs = fcm.fuzzy_membership

>>> act_pvh = fcm.fuzzy_implication(mfs['+VH'], weight= 0.33, method ='Mamdani')
>>> act_pm = fcm.fuzzy_implication(mfs['+M'], weight=0.16, method ='Mamdani')
>>> act_ph = fcm.fuzzy_implication(mfs['+H'], weight=0.5, method ='Mamdani')

activatedMamdani = {'+vh' : act_pvh, '+h' : act_ph, '+m' : act_pm}
\end{lstlisting}

\subsubsection{Step 3: Aggregate fuzzy membership functions}
In the third step, we must aggregate the activated membership functions taken from the previous step. This is commonly done by applying the family maximum aggregation operation. Alternative methods for aggregating membership functions include the family Algebraic Sum \eqref{eq:algSum}, the family Einstein Sum \eqref{eq:eSum} and the family Hamacher Sum \eqref{eq:hSum} \cite{Piegat:2001}.

\begin{equation}\label{eq:algSum}
    f(x, y)=x + y - x \times y
\end{equation}

\begin{equation} \label{eq:eSum}
 f(x, y) = \frac{(x + y)}{(1 + x \times y)}
\end{equation}

\begin{equation} \label{eq:hSum}
f(x, y) = \frac{(x + y - 2 \times x \times y)}{(1 - x \times y)}  
\end{equation}

One can use the {\ttfamily aggregate()} method to aggregate the activated membership functions (see the available aggregation method in Table~\ref{tab:aggregation} and the code snippet below).

\begin{lstlisting}[language=Python]
>>> import functools

>>> aggregated = functools.reduce(lambda x,y: 
                    fcm.aggregate(x=x, y=y, method='fMax'),
                    [activatedMamdani[i] for i in activatedMamdani.keys()])
\end{lstlisting}

\begin{table}[t!]
\centering
\begin{tabular}{lllp{7.4cm}}
\hline
Argument         & Option & Description \\ \hline
{\ttfamily method}        & `fMax'  & Family maximum \\
                & `algSum'  &  Family Algebraic Sum\\
                 & `eSum'  & Family Einstein Sum \\
                 & `hSum'  & Family Hamacher Sum \\\hline
\end{tabular}
\caption{\label{tab:aggregation} Aggregation rules}
\end{table}

where x and y are the membership values of the linguistic terms involved in the problem domain after the application of the implication rule presented in the previous step.

\subsubsection{Step 4: Defuzzify the aggregated membership functions}

The last step includes the calculation of the crisp value based on the aggregated membership functions (a.k.a. defuzzification). Among the available defuzzification methods (see the available defuzzification methods in Table~\ref{tab:defuzz}) the most commonly used method is the centroid method (a.k.a. center of gravity) \cite{Stach:2005}.

\begin{figure}[H]
\centering
\includegraphics[width = \textwidth]{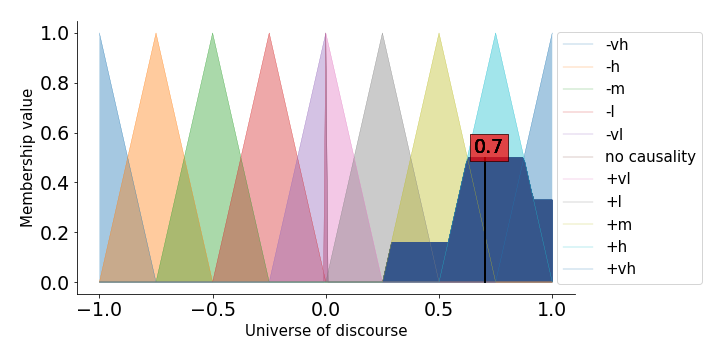}
\caption{\label{fig:defuzz} Defuzzification of the aggregated membership functions}
\end{figure}

We can apply the dedicated {\ttfamily defuzz()} method to derive the crisp value (see Figure \ref{fig:defuzz} and the code snippet below).

\begin{lstlisting}[language=Python]
>>> dfuz = fcm.defuzz(x=fcm.universe, mfx=aggregated, method='centroid')
\end{lstlisting}

\begin{table}[t!]
\centering
\begin{tabular}{lllp{7.4cm}}
\hline
Argument         & Option & Description \\ \hline
{\ttfamily method}        & `centroid'  & Centroid \\
                & `bisector'  &  Bisector\\
                 & `mom'  & Mean of maximum \\
                 & `som'  & Min of max \\
                 & `lom'  & Max of maximum \\\hline
\end{tabular}
\caption{\label{tab:defuzz} Defuzzification methods}
\end{table}

The above mentioned four steps can either be done and controlled independently as we have shown, or users can rely on a single {\ttfamily build()} method that implements those steps to calculate the numerical weights for all the concept pairs in the data (see the code snippet below). The method returns a pandas dataframe with the calculated weights.

\begin{lstlisting}[language=Python]
>>> data = fcm.read_data(file_path) 
>>> weight_matrix = fcm.build(data=data, implication_method = Mamdani, 
                    aggregation_method = 'fMax', defuzz_method='centroid')
\end{lstlisting}

\begin{lstlisting}[language=Python]
    C1      C2	  C3      C4	
C1 0.000000 0.703218 0.000000 0.000000
C4 0.000000 0.000000 0.000000 0.000000
C2 0.608308 0.000000 0.000000 0.000000
C3 0.555732 0.000000 0.000000 0.159091
\end{lstlisting}

\section{Simulating the system behavior with FCMs} \label{sec:simFCM}
The dynamics of the specified FCM are examined by simulating its behavior over discrete simulation steps. In each simulation step, the concept values are updated according to a defined inference method \cite{Papageorgiou2011}. The \textit{Simulator} module implements the following three types of inference methods (see the available options in Table~\ref{tab:inftrans}):

\begin{itemize}
  \item Kosko:
    \begin{equation} \label{eq:Kosko}
    A_{i}^{t+1}=f(\sum_{j=1}^n A_{j}^t * W_{ji}) 
    \end{equation}
  \item Modified Kosko:
    \begin{equation} \label{eq:MKosko}
    A_{i}^{t+1}= f(A_{i}+\sum_{j=1}^n A_j^t * W_{ji})
    \end{equation}
  \item Rescaled:
    \begin{equation} \label{eq:Rescaled}
    A_{i}^{t+1} = f((2A_{i} -1)  +\sum_{j=1}^n (2A_{j}^t -1) * W_{ji}) 
    \end{equation}
\end{itemize}

where $A_{j}^t$ is the value of concept $j$ at the simulation step $t$ and $W_{j,i}$ is the causal impact of concept $j$ on concept $i$. Note that a (transfer) function $f(x)$ is applied to the result. As shown in the equations above, this function is necessary to keep values within a certain range (e.g., [0,1] for sigmoid function or [-1,1] for hyperbolic tangent). In the current version, four such functions are implemented (see the available options in Table~\ref{tab:inftrans}):

\begin{itemize}
  \item Sigmoid:
    \begin{equation} \label{eq:sigmoid}
    f(x)=\frac{1}{1+e^{-\lambda x}}, x\in \mathbb{R}; \text{binds node values to}\; [0,1]
    \end{equation}
  \item Hyperbolic tangent:
    \begin{equation} \label{eq:tanh}
    f(x)=tanh(x)=\frac{sinh(x)}{cosh(x)}=\frac{e^{2x}-1}{e^{2x}+1}, x\in\ \mathbb{R}; \text{binds node values to}\; [-1,1]
    \end{equation}
  \item Bivalent:
    \begin{equation} \label{eq:bi}
    f(x)=
    \begin{cases}
      1, & x > 0\\
      0, & x\leq 0
    \end{cases}, x\in\ \mathbb{R}; \text{binds node values to}~\{0,1\}
    \end{equation}
    \item Trivalent:
    \begin{equation} \label{eq:tri}
    f(x)=
    \begin{cases}
      1 & x > 0\\
      0 & x= 0\\
      -1 & x<0
    \end{cases},  x\in\ \mathbb{R};  \text{binds node values to}~\{-1, 0, 1\} 
    \end{equation}
\end{itemize}

where, x is the defuzzified value and the $\lambda$ is a steepness parameter for the sigmoid function.

\begin{table}[t!]
\centering
\begin{tabular}{lllp{7.4cm}}
\hline
Argument         & Option & Description \\ \hline
{\ttfamily inference}        & `kosko'  & Kosko \\
                & `mKosko'  &  Modified Kosko\\
                 & `rescaled'  & Rescaled \\ \hline
{\ttfamily transfer}    & `sigmoid'  & Sigmoid \\
                    & `tanh'  & Hyperbolic tangent \\
                    & `bivalent'  & Bivalent \\
                    & `trivalent'  & Trivalent \\\hline
\end{tabular}
\caption{\label{tab:inftrans} Inference and transfer methods}
\end{table}

The simulation is run until either of two conditions is met: (1) some concepts of interest have a difference lower than a given threshold between two consecutive steps, or (2) a user-defined maximum number of iterations is reached. If we denote by S the subset of concepts of interest (i.e., the outputs of the FCMs), then the first condition can be stated as:
\begin{equation}
\exists t \in {1, 2, \ldots, T-1} : |A^{t+1} - A^t| < threshold
\end{equation}

\begin{figure}[H]
\centering
\includegraphics[width = \textwidth]{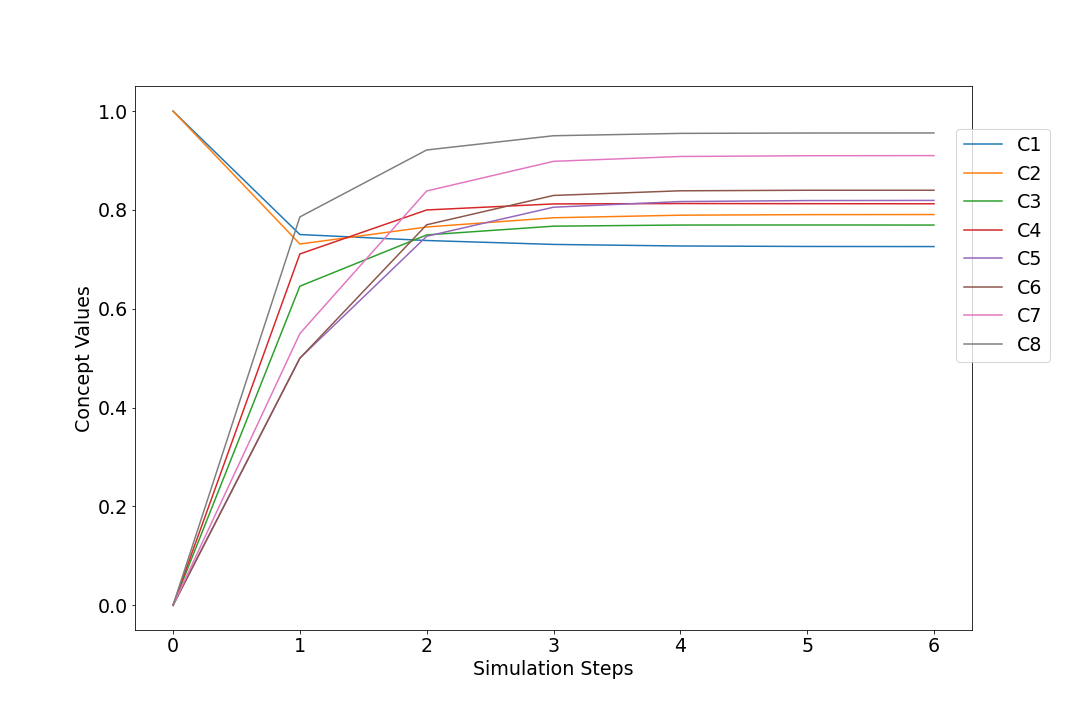}
\caption{\label{fig:simulations} Simulation results of 8 concepts}
\end{figure}

The {\ttfamily simulate()} method takes the initial state vector and the FCM weight matrix (a.k.a., connection matrix) and applies one of the mentioned update functions over number of simulation steps (see the simulation results in Figure \ref{fig:simulations}). One can specify the output concepts by supplying a list of these concepts to the respective {\ttfamily $output\_concepts$} argument. If the {\ttfamily $output\_concept$} argument is not specified then all the concepts in the FCM are treated as output concepts and the simulation stops when all the concepts change by less than the threshold between two consecutive steps.

\begin{lstlisting}[language=Python]
>>> import pandas as pd
>>> from fcmpy import FcmSimulator

>>> w = np.asarray([[0.0, 0.0, 0.6, 0.9, 0.0, 0.0, 0.0, 0.8],
                [0.1, 0.0, 0.0, 0.0, 0.0, 0.0, 0.2, 0.5],
                [0.0, 0.7, 0.0, 0.0, 0.9, 0.0, 0.4, 0.1],
                [0.4, 0.0, 0.0, 0.0, 0.0, 0.9, 0.0, 0.0],
                [0.0, 0.0, 0.0, 0.0, 0.0, -0.9, 0.0, 0.3],
                [-0.3, 0.0, 0.0, 0.0, 0.0, 0.0, 0.0, 0.0],
                [0.0, 0.0, 0.0, 0.0, 0.0, 0.8, 0.4, 0.9],
                [0.1, 0.0, 0.0, 0.0, 0.0, 0.1, 0.6, 0.0]])

>>> weight_matrix = pd.DataFrame(w, 
                    columns=['C1','C2','C3','C4','C5','C6','C7','C8'])

>>> init_state = {'C1': 1, 'C2': 1, 'C3': 0, 'C4': 0, 'C5': 0,
                    'C6': 0, 'C7': 0, 'C8': 0}

>>> sim = FcmSimulator()

>>> res = sim.simulate(initial_state=init_state, weight_matrix=weight_matrix,
        transfer='sigmoid', inference='mKosko', 
        thresh=0.001, iterations=50)
\end{lstlisting}

\begin{lstlisting}[language=Python]
The values converged in the 7 state (e <= 0.001)

        C1        C2        C3        C4        C5        C6        C7        C8
0  1.000000  1.000000  0.000000  0.000000  0.000000  0.000000  0.000000  0.000000
1  0.750260  0.731059  0.645656  0.710950  0.500000  0.500000  0.549834  0.785835
2  0.738141  0.765490  0.749475  0.799982  0.746700  0.769999  0.838315  0.921361
3  0.730236  0.784168  0.767163  0.812191  0.805531  0.829309  0.898379  0.950172
4  0.727059  0.789378  0.769467  0.812967  0.816974  0.838759  0.908173  0.954927
5  0.726125  0.790510  0.769538  0.812650  0.818986  0.839860  0.909707  0.955666
6  0.725885  0.790706  0.769451  0.812473  0.819294  0.839901  0.909940  0.955774
\end{lstlisting}

\section{Machine learning for FCMs} \label{sec:MlFCM}
%%%%%%
As shown in the previous sections, FCMs are often constructed based on experts' knowledge about the system. In certain domains of applications, modelers either optimize the FCMs constructed by the experts and/or constructing FCMs entirely based on the data collected about the systems. A set of machine learning algorithms were developed to meet these tasks and have previously been applied to numerous fields such as the optimization of industrial processes \cite{papageorgiou2006unsupervised, stach2008data, papageorgiou2011learning}, decision making \cite{poczeta2020fuzzy}, and classification \cite{napoles2014two,napoles2021pattern}. In the proposed library, we include three types of algorithms used for edge optimization, FCM generation, and classification. We used the state-of-the-art methods~\cite{napoles2021pattern, napoles2020deterministic} and the foundational ones that have been widely adopted~\cite{papageorgiou2006unsupervised, stach2010learning}.

% widely used as well as the ones that were recently developed by the scientists in the field. 
\subsection{Hebbian learning}
\label{sec:hebbian_learning}
One of the weaknesses of an FCM constructed by the experts is its potential convergence to undesired regions. For example, given an intervention scenario, the model may predict only extreme values such as 0 or 1~\cite{lavin2018should}. To overcome this weakness~\cite{papageorgiou2006unsupervised} proposed two learning strategies, namely the Active Hebbian Learning (AHL) and the Non-Linear Hebbian Learning (NHL) algorithms that are based on the Hebbian learning rule. The task of the proposed algorithms is to modify the initial FCM connection matrix constructed by the expert such that the chosen nodes (called Desired Output Concepts \textit{DOCs}) always converge within the desired range. Both algorithms are similar to FCM simulation, with the main difference being that concepts' values \textit{and weights'} are updated at each time step, whereas during a simulation, only the concepts values are changing. 

In the NHL algorithm, all nodes ($A_i$) and weights ($W_{ij}$: a direct edge from node $i$ to $j$) values are \textit{simultaneously} updated at each time step. In AHL, nodes and weights are updated \textit{asynchronously} based on a sequence of activation patterns specified by the user. During each simulation time step, a new node becomes an \textit{``activated node''}; only this node and its incoming edges are updated, while everything else remains unchanged.  Along with optimizing existing edges, AHL creates new connections between the concepts, which may be an undesirable behavior if the modeler's intent is to tweak the weights rather than create connections that have not been endorsed by experts. 

% \begin{equation}
% \centering
% A_{i}^{k+1} = f(A_{i}^{k} +\sum_{j=1 \atop j\neq i}^{N}A_{j}^{k}W_{ji}^{k})\label{eq:nodeHL}
% \end{equation}

% \begin{equation}
% \centering
% W_{ij}^{k+1} = \gamma W_{ij}^{k} + \eta A_{i}^{k} (A_{j}^{k} - sgn(W_{ij}) W_{ij}^{k}A_{i}^{k})\label{eq:weightNHL}
% \end{equation}

% \begin{equation}
% \centering
% W_{ij}^{k+1} = (1 - \gamma)W_{ij}^{k} + \eta A_{i}^{act,k} (A_{j}^{k}-W_{ij}^{k}(A_{i}^{act,k}))\label{eq:weightAHL}
% \end{equation}

% \begin{equation}
% \centering
% A_{i}^{act,k+1} = f(A_{i}^{act,k} +\sum_{j=1 \atop j\neq i}^{N}A_{j}^{k}W_{ji}^{k})\label{eq:nodeAHL}
% \end{equation}

The learning process continues until two termination conditions are fulfilled. First, the fitness function ($F_1$) is calculated for each DOC per equation~\eqref{eq:termNHL1}. If value of $F_1$ for each DOC declines at each time step, and the DOCs values are within a desired range, the first termination condition is fulfilled. Second, it is crucial to determine whether the values of the DOCs are stable, i.e. if their values vary with each step more than a threshold $e$ shown in Equation~\ref{eq:termNHL2}. This threshold should be determined experimentally, and it is recommended to be set between $0.001$ and $0.005$~\cite{papageorgiou2006unsupervised}. If the change is lower than the threshold, the second termination condition is fulfilled.
%Equations \eqref{eq:avgDOC},\eqref{eq:termNHL1} and \eqref{eq:termNHL2} show $T_{j}$, first and second termination condition respectively.  

% \begin{equation}
% \centering
% T_{j} = \frac{T_{j}^{min}-T_{j}^{max}}{2}\label{eq:avgDOC}
% \end{equation}

\begin{equation}
\centering
F_{1} = \sqrt{ | DOC_j^k - \frac{DOC_j^{min} - DOC_j^{max}}{2} |^2 }\label{eq:termNHL1}
\end{equation}

\begin{equation}
\centering
F_{2} = |DOC_{j}^{k+1}-DOC_{j}^{k}| < e\label{eq:termNHL2}
\end{equation}

If the termination conditions are satisfied then the learning process may stop, otherwise, it will continue until a maximum number of steps is reached (we set the default value to 100).
In order to use these methods, the user has to provide the initial weight matrix, initial concept values, and the DOCs. In addition, these variables are necessary, in most cases, algorithms converge only for a specific combination of values of the hyperparameters: learning rate ($\eta$), decay coefficient ($\gamma$), and slope of sigmoid function. The sample values used in several case studies are slope $[0.9,1.01]$, decay for NHL $[0.99,1.0]$, decay for AHL $[0.01,0.1]$ and learning rate $[0.001,0.1]$. The optimization of an FCM from a water tank case study \cite{papageorgiou2004active,ren2012learning,papakostas2011training} using the algorithms above is demonstrated in the code snippet below.

\begin{lstlisting}[language=Python]
>>> from fcmpy import NHL
>>> import numpy as np

# initial values of weight matrix
>>> w_init = np.asarray([[0,-0.4,-0.25,0,0.3],
                        [0.36,0,0,0,0],
                        [0.45,0,0,0,0],
                        [-0.9,0,0,0,0],
                        [0,0.6,0,0.3,0]])

>>> w_init = pd.DataFrame(w_init, 
                            columns=['C1', 'C2', 'C3', 'C4', 'C5'],
                            index = ['C1', 'C2', 'C3', 'C4', 'C5'])

# initial values of the concepts
>>> init_states = {'C1' : 0.40, 'C2': 0.7077, 
                    'C3': 0.612, 'C4': 0.717, 'C5': 0.30}

# DOCs 
>>> doc_values = {'C1':[0.68,0.74], 'C5':[0.74,0.8]}

# NHL
>>> nhl = NHL(state_vector=init_states, weight_matrix=w_init,         
                doc_values=doc_values)
>>> res = nhl.run(learning_rate = 0.01, l=.98, iterations=100)
\end{lstlisting}

\begin{lstlisting}[language=Python]
The NHL learning process converged at step 63 with the 
learning rate eta = 0.01 and decay = 1!
    C1        C2        C3        C4        C5
C1  0.000000 -0.200310 -0.023806  0.000000  0.472687
C2  0.539068  0.000000  0.000000  0.000000  0.000000
C3  0.571531  0.000000  0.000000  0.000000  0.000000
C4 -0.832174  0.000000  0.000000  0.000000  0.000000
C5  0.000000  0.710523  0.000000  0.496934  0.000000
\end{lstlisting}

The {\ttfamily AHL.run()} method has an additional {\ttfamily auto\_learn} argument; if set to {\ttfamily True}, then {\ttfamily True} then the algorithm automatically updates the hyperparameters during the learning process.

\begin{lstlisting}[language=Python]
# AHL
>>> activation_pattern = {0:['C1'], 1:['C2', 'C3'], 2: ['C5'], 3: ['C4']}

>>> ahl = AHL(state_vector=init_states, weight_matrix=w_init,  
                activation_pattern=activation_pattern,
                doc_values=doc_values)
>>> res = ahl.run(decay=0.03, learning_rate = 0.01, l=1, iterations=100,
                    transfer= 'sigmoid', thresh = 0.002, auto_learn=False,
                    b1=0.003, lbd1=0.1, b2=0.005, lbd2=1)
\end{lstlisting}

\begin{lstlisting}[language=Python]
The AHL learning process converged at step 19 with 
the learning rate eta = 0.01 and decay = 0.03!

        C1        C2        C3        C4        C5
C1  0.000000 -0.128532 -0.060395  0.071200  0.218170
C2  0.245859  0.000000  0.068981  0.076592  0.074289
C3  0.288257  0.069457  0.000000  0.070342  0.068190
C4 -0.386349  0.073807  0.067187  0.000000  0.073991
C5  0.070113  0.368913  0.069145  0.223312  0.000000
\end{lstlisting}

\begin{figure}[ht!]
\centering
  \subfloat[Initial weight matrix]{\includegraphics[width=0.5\textwidth]{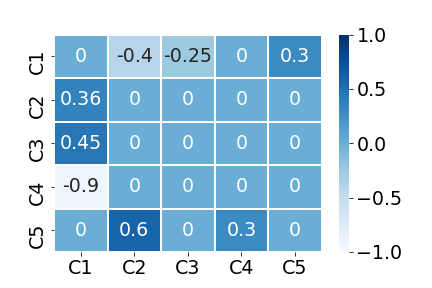}\label{fig:Winit}}
  \hfill
  \subfloat[AHL]{\includegraphics[width=0.5\textwidth]{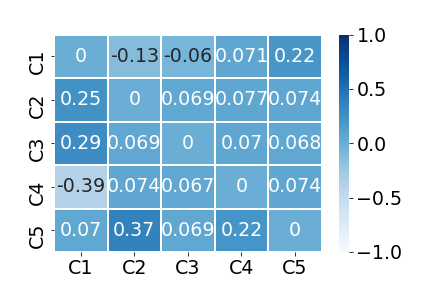}\label{fig:WAhl}}
    \hfill
  \subfloat[NHL]{\includegraphics[width=0.5\textwidth]{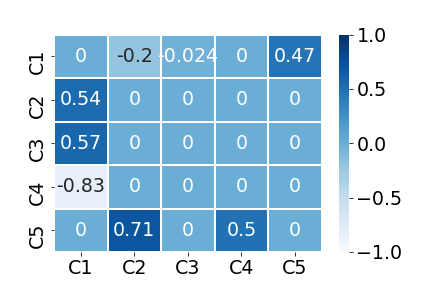}\label{fig:WNhl}}
  \caption{\label{fig:Hebbian} a) initial weight matrix, b) weight matrix optimized by AHL, and c) weight matrix optimized by NHL.}
\end{figure}

If the learning process was successful, the {\ttfamily run} method will return the optimized weight matrix as a dataframe. Successful outputs of NHL Figure~\ref{fig:WNhl} and AHL Figure~\ref{fig:WAhl} algorithm. 

\subsection{Real-coded genetic algorithm (RCGA)}
In certain domains of application, one has longitudinal data about the state variables included in the FCM and wants to find an FCM connection matrix that generates data that is close enough to the collected data~\cite{khan2004fuzzy, poczketa2015learning}. In this regards,~\cite{stach2010learning} proposed a real coded genetic algorithm for searching for an optimal FCM connection matrix. The proposed algorithm, named RCGA, builds on genetic algorithms as follows.

\begin{figure}[H]
\centering
\includegraphics[width = \textwidth]{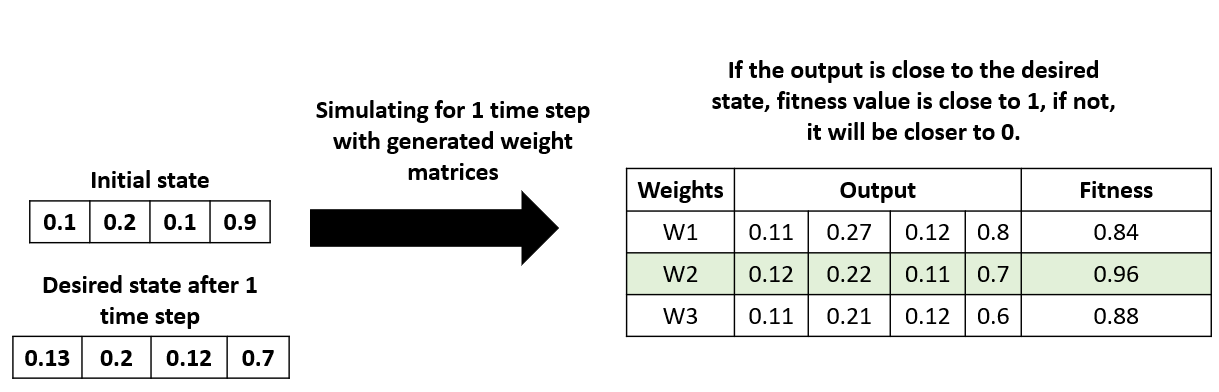}
\caption{\label{fig:rcgafit} Example of fitness values calculation.}
\end{figure}

The search process of the RCGA includes the following six steps: 1) initialization, 2) evaluation, 3) selection, 4) recombination, 5) mutation, and 6) replacement. In the initialization step, the algorithm generates a population of random solutions, that is a set of random weight matrices. Each solution is an $N~\times~N$ connection matrix. In the evaluation step, each candidate solution in the population is evaluated based on a fitness function shown in Equation~\eqref{eq:fitness} and Table~\ref{tab:rcga}. We can observe how the fitness function is calculated on a simple example, shown in Figure~\ref{fig:rcgafit}).

\begin{equation}
\centering
\begin{split}
    Error & = \alpha \sum\limits_{t=1}^{T-1} \sum\limits_{n=1}^{N-1} |C_n(t) - \hat{C}_n(t)|^p \\
    & Fitness = \frac{1}{a*Error + 1}
\end{split}
\label{eq:fitness}
\end{equation}

\begin{table}[h!]
\centering
\begin{tabular}{lllp{7.4cm}}
\hline
\textbf{Variable}                    & \textbf{Description}                                        \\ \hline
$\alpha$       & Normalization parameter                            \\ \hline
T                           & Number of chromosomes (elements in the generation) \\ \hline
N                           & Number of variables in the chromosome              \\ \hline
C(t)                        & Observed data at time t                                    \\ \hline
$\hat{C}(t)$ & State vector at step t                   \\ \hline
p                           & Defines the type of norm (default 1)           \\ \hline
a                           & Defines the type of norm (default 100)        \\ \hline
\end{tabular}
\caption{Variables of fitness function used in RCGA algorithm.}
\label{tab:rcga}
\end{table}
In the selection step, candidate solutions are selected for mating (a.k.a., recombination). In each step, the algorithm randomly selects between two selection mechanisms (roulette wheel and tournament selection strategies). For the recombination step, the algorithm implements the recommended one point crossover operation with a probability of crossover specified by the user ({\ttfamily p\_recombination}) \cite{stach2010learning}. The crossover operation creates new solutions based on the solutions selected in the previous step. Next, the algorithm decides whether the new solutions produced in the previous step should undergo mutations. In the mutation step, the algorithm chooses between random and non-uniform mutation operations with a probability defined by the user ({\ttfamily p\_mutation}). The replacement step is determined by the evolutionary approach specified by the user. The algorithm proposed by \cite{stach2010learning} is based on a generational approach where in each step the new generation of solutions replace the old generation. Alternatively, the user could choose a steady state approach (a.k.a., SSGA) where in each step only two new solutions are produced and a decision is made whether the new chromosomes should be inserted back into the population. The current implementation of the SSGA uses a replacement strategy based on the concept of useful diversity (described in depth in \cite{lozano2008}). To use the {\ttfamily RCGA} module, one needs to initialize the {\ttfamily RCGA} class by specifying the longitudinal data about the system, the population size, and the genetic approach to use (i.e., generational or steady state). The additional parameters that can be modified by the user are presented in Table~\ref{tab:rcgaParam}. Other parameters that can be modified by the user can be found in the documentation of the package available on PyPI.
\begin{table}[H]
\centering
\begin{tabular}{lllp{7.4cm}}
\hline
Argument         & Option & Description \\ \hline
normalization\_type  & `L1'  & L1 normalization \\
                & `L2'  &  L2 normalization\\
                 & `LInf'  & L infinity normalization \\ \hline

inference   & `kosko'  & Kosko \\
                    & `mKosko'  & Modified kosko \\
                    & `rescaled'  & Rescaled inference method \\ \hline 
                    
transfer    & `sigmoid'  & Sigmoid \\
                    & `bivalent'  & Bivalent \\
                    & `trivalent'  & Trivalent \\
                 & `tanh'  & Hyperbolic tangent \\\hline
\end{tabular}
\caption{\label{tab:rcgaParam} Inference and transfer methods}
\end{table}
The output of the learning process is the weight matrix with the highest fitness value throughout the search process. An example of generating FCM by the RCGA using historical data on water tank case study~\cite{papageorgiou2004active} presented in Section \ref{sec:hebbian_learning} is demonstrated in the code snippet below. We give the user an option to choose: {\ttfamily $population\_size$} which is a number of weights matrices generated at each time step and {\ttfamily threshold} that is a minimum fitness value of at least one weight matrix in a generation, for the algorithm to succeed. To ensure the user does not get stuck in an infinite loop, we define {\ttfamily $n\_iterations$} after which algorithm will terminate if the max fitness function of the $n\_iterations~-~1$ generation was less than a {\ttfamily threshold}.

% \begin{figure}[H]
% \centering
% \includegraphics[width=150mm]{rcga.PNG}
% \caption{\label{fig:rcga} Flow chart of RCGA algorithm.}
% \end{figure}
% check if for sure

\begin{lstlisting}[language=Python]
# Generate Longitudinal Data
>>> sim = FcmSimulator()
>>> dataWaterTank = sim.simulate(initial_state=init_states,
                                weight_matrix=w_init, transfer='sigmoid', 
                                inference='mKosko', thresh=0.001,
                                iterations=50, l=1)
# Select two time points
>>> dataWaterTank = dataWaterTank.iloc[:3]

# Generational Approach
>>> rcga = RCGA(data=dataWaterTank, population_size=100,
                ga_type='generational')
>>> rcga.run(n_iterations=30000, threshold=0.99)

# Steady State Approach
>>> rcga = RCGA(data=dataWaterTank, population_size=100,
                ga_type='ssga')
>>> rcga.run(n_iterations=30000, threshold=0.99)
\end{lstlisting}

The RCGA solution and the associated fitness score can be accessed in the {\ttfamily rcga.solution} and {\ttfamily rcga.fitness} fields.

\begin{lstlisting}[language=Python]
>>> rcga.fitness
\end{lstlisting}

\begin{lstlisting}[language=Python]
0.9790443073348711
\end{lstlisting}

\begin{lstlisting}[language=Python]
>>> rcga.solution
\end{lstlisting}

\begin{lstlisting}[language=Python]
        C1        C2        C3        C4        C5
C1   0.851948   0.550687  0.045434  0.648337  0.077349
C2  -0.870796  -0.944261 -0.684952 -0.829811  0.660863
C3   0.807396  -0.075391  0.223011  0.299976 -0.442272
C4  -0.479566   0.724956  0.048086  0.352487  0.120083
C5  -0.030473   0.068480  0.330551  0.003072 -0.347919
\end{lstlisting}

The learned FCM connection matrix can be validated by calculating the in-sample and out-sample errors by using the dedicated {\ttfamily ISE} and {\ttfamily OSE} modules (see the code snippet below) \cite{stach2010learning}.

\begin{lstlisting}[language=Python]
>>> from fcmpy import ISE
>>> from fcmpy import OSE

>>> val_ise = ISE()
>>> val_ose = OSE()
>>> error_ise = val_ise.validate(initial_state=init_state_WT,
                weight_matrix=rcga.solution, data=dataWaterTank, 
                transfer='sigmoid', inference='mKosko', l=1)

>>> error_ose, std = val_ose.validate(weight_matrix=rcga.solution,
                                    data=dataWaterTank, low=0, high=1, 
                                    k_validation=100, transfer='sigmoid', 
                                    inference='mKosko', l=1)
                                    
>>> print(f"in-sample error: {error_ise}, \
            out-of-sample error: {error_ose}, std: {std}" )
\end{lstlisting}

\begin{lstlisting}[language=Python]
in-sample error: 0.0129, out-of-sample error: 0.0891, std: 0.0669
\end{lstlisting}

\subsection{Classification algorithms}
% In addition to classification, one of the proposed algorithms reveals which features are the key factors in decision making. 
It is attractive for the researchers to choose FCMs for classification tasks, over other popular tools such as neural networks. FCMs are easily explainable, which is a great advantage over \textit{black box models} and, in many cases, equally accurate~\cite{napoles2021pattern, napoles2020deterministic}. We give user a choice of two methods: Evolving Long-term Cognitive 
Networks (ELTCN)~\cite{napoles2021pattern} and deterministic learning (LTCN-MP)\footnote{LSTCNs are a variant of FCMS where weights are nor expected to be in the [-1,1] interval or have a causal meaning. We use regularization in order to keep them within that range.}~\cite{napoles2020deterministic}. 

LTCN-MP and ELTCN use the same topology (a fully connected FCM containing features nodes and class nodes) but the former produces numerical outputs (suitable for regression) while the latter produces nominal outputs (suitable for classification). In LTCN-MP and ELTCN algorithms (i) input variables are located in the inner layer and output variables in the outer layer, (ii) weights connecting the inputs are computed in an unsupervised way by solving a least squared problem, and (iii) weights connecting inputs with outputs are computed using the Moore-Penrose pseudo-inverse. Overall, we can say that LTCN-MP and ELTCN use the same topology but the former produces numerical outputs while the latter produces nominal outputs (decision classes).

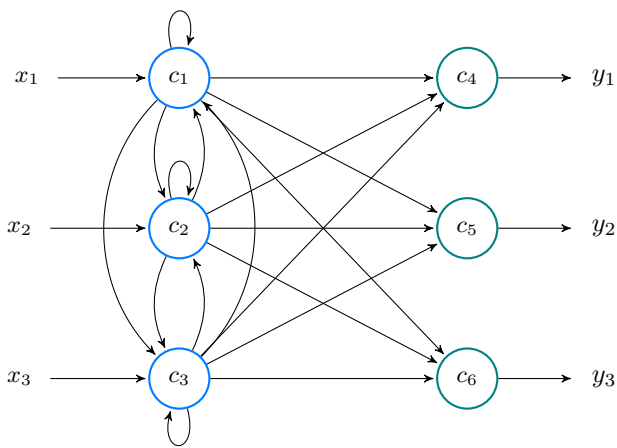
\begin{figure}[!ht]
	\centering
	\begin{tikzpicture}[->,>=stealth',shorten >=1pt,auto,node distance=2cm,main node/.style={minimum size=8mm,circle,draw=mycolor,font=\sffamily\bfseries}]
	\node[main node] (1) [font=\small, thick] {$c_1$};
	\node[main node] (2) [font=\small, thick, below of=1] {$c_2$};
	\node[main node] (3) [font=\small, thick, below of=2] {$c_3$};
	\node[main node] (4) [font=\small, thick, draw=teal, right = 3cm of 1] {$c_4$};
	\node[main node] (5) [font=\small, thick, draw=teal, right = 3cm of 2] {$c_5$};
	\node[main node] (6) [font=\small, thick, draw=teal, right = 3cm of 3] {$c_6$};
	\node[main node] (7) [font=\small, draw=none, left = 1.2cm of 1] {$x_1$};
	\node[main node] (8) [font=\small, draw=none, left = 1.3cm of 2] {$x_2$};
	\node[main node] (9) [font=\small, draw=none, left = 1.3cm of 3] {$x_3$};
	\node[main node] (10) [draw=none, right = 1cm of 4] {$y_1$};
	\node[main node] (11) [draw=none, right = 1cm of 5] {$y_2$};
	\node[main node] (12) [draw=none, right = 1cm of 6] {$y_3$};
	\path[every node/.style={ font=\sffamily}]
	(1) 
	edge [bend right=25] node [font=\normalsize, left] {} (2)
	edge [bend right=45] node [font=\normalsize, pos=0.76, left] {} (3)
	edge node [font=\normalsize, pos=0.8, sloped, above] {} (4)
	edge node [font=\normalsize, pos=0.8, sloped, above] {} (5)
	edge node [font=\normalsize, pos=0.8, sloped, above] {} (6)
	edge [loop above] node {} (1)
	(2) 
	edge [bend right=25] node [font=\normalsize, left] {} (3)
	edge [bend right=25] node [font=\normalsize, left] {} (1)
	edge node [font=\normalsize, pos=0.8, sloped, above] {} (4)
	edge node [font=\normalsize, pos=0.8, sloped, above] {} (5)
	edge node [font=\normalsize, pos=0.8, sloped, above] {} (6)
	edge [loop above] node {} (2)
	(3) 
	edge [bend right=25] node [font=\normalsize, right] {} (2)
	edge [bend right=45] node [font=\normalsize, pos=0.76, left] {} (1)
	edge node [font=\normalsize, pos=0.8, sloped, above] {} (4)
	edge node [font=\normalsize, pos=0.8, sloped, above] {} (5)
	edge node [font=\normalsize, pos=0.8, sloped, above] {} (6)
	edge [loop below] node {} (3)
	(7) edge node [above] {} (1)
	(8) edge node [above] {} (2)
	(9) edge node [above] {} (3)
	(4) edge node [above] {} (10)
	(5) edge node [above] {} (11)
	(6) edge node [above] {} (12);
	\end{tikzpicture}
	\caption{Neural model comprised of $M=3$ input neurons ($c_1$, $c_2$, $c_3$) and $N=3$ output neurons ($c_4$, $c_5$, $c_6$). Overall, the network has $P=M+N=6$ neurons. In this example, the input signal is denoted as $x_i$ while output signal is denoted as $y_i$. These two values correspond to $a_i^{(0)}$ and $a_i^{(T)}$, respectively.}
	\label{fig:network}
\end{figure}

An example of a model's structure with three features and three classes is shown in Figure~\ref{fig:network}.

% \begin{figure}[H]
% \centering
% \includegraphics[width=50mm]{eltnc.PNG}
% \caption{\label{fig:eltcn} Structure of a model generated by ELTCN algorithm.}
% \end{figure}
%% Please remove the unused imports

The user has to provide the path to the directory where the data file (\textit{.arff} format) is located\footnote{Currently, these two algorithms only accept \textit{.arff} files, we are planning to accommodate more data files formats in the future versions.}. It is necessary that values of the features are normalized in the range between 0 and 1. Multiple data sets can be utilized and the results of each of them is saved in a dictionary, using the filename as a key. After running the learning process, the output consist of \textit{k} weight matrices, where \textit{k} is a number of validation folds (default 5) and the weight matrix of the connections between classes and feature nodes. We also provide users with automatically generated histograms showing values of the loss function and weights for each fold. In our examples we are using the \textit{Iris data set}, a popular data set containing various measurements of iris flowers, such as sepal or petal length~\cite{fisher1936use}.

\begin{lstlisting}[language=Python]
>>> from fcmpy.ml.classification.eltcn import run
>>> path = 'data'
>>> results = run(path)

# average of the feature weight matrices 
>>> print(results['irisnorm.arff']['avgW'])

# weight matrix connecting feature nodes with class nodes 
>>> print(results['irisnorm.arff']['classW'])
\end{lstlisting}

\begin{lstlisting}[language=Python]
[[ 0.4382333 , -0.09812646,  0.79643613,  0.97163904],
[-0.2016794 ,  0.2925336 , -0.21699643,  0.12452463],
[-0.0038293 , -0.08575615,  0.47770625,  0.52030253],
[ 0.24917993,  0.21717176,  0.30113676,  0.34600133]]

[[-0.9135318 ,  0.15331155,  0.24873467],
[-1.        ,  0.42972276,  0.17217618],
[-0.5610481 , -0.58390266,  0.79575956],
[-0.60481995, -0.6718271 ,  0.72851056]]
\end{lstlisting}

LSTCN-MP focuses on discovering which and how features of the data set are important for the classification task as well as finding the weights connecting the inputs and outputs. Next, the LSTCN-MP algorithm outputs a 1-D array with values in the [-1,1] range. The absolute values represent how important the features are for the classification task. In order to use the algorithm, the user has to provide the path of data sets as list under a key \textit{sources} and then use that dictionary as an input to LSTCN-MP algorithm.

\begin{lstlisting}[language=Python]
    >>> sources = [''datasets/iris.arff'',``datasets/pima.arff'']
    >>> params = {'sources':sources}
\end{lstlisting}

 Various hyperparameters can be used. Their default values and description can be found in~\cite{napoles2020deterministic}, our library documentation, or Table~\ref{tab:mp}.

% \usepackage{booktabs}
\begin{table}[ht!]
\centering
\begin{tabular}{@{}llll@{}}
\hline
Keys        & Values          & Description     & Default value                                     \\\hline
``M''         & \textit{int}    & Output variables & 1                         \\\hline
``T''         & \textit{int}    & FCM Iterations   & 1                                    \\\hline
``folds''     & \textit{int}    & Number of folds in a (stratified) K-Fold &10 \\\hline
``output''    & \textit{string} & Output csv file  &``./output.csv''             \\\hline
``p''         & \textit{array}  & parameters of \textit{logit} and \textit{expit} functions &$\{1.0, 1.0, 1.0, 1.0\}$               \\\hline
``sources''   & \textit{array}  & array with path of the dataset files & None                 \\\hline
``verbosity'' & \textit{bool}   & verbosity &False                           \\\hline
\end{tabular}
\caption{\label{tab:mp} Keys and values of the input dictionary to the deterministic algorithm.}
\end{table}

\begin{lstlisting}[language=Python]
>>> import fcmpy.ml.classification.FCM_MP as mp
>>> import matplotlib.pylab as plt
>>> sources = ['iris.arff']
>>> params = {'sources':sources}
>>> out = mp.run(**params)
# feature importance for classification purposes
>>> fig, ax = plt.subplots()
>>> ax.bar(range(len(out[0]['importance'].flatten())),
    height=out[0]['importance'].flatten())
# connections between features and class nodes 
>> print(out['weights'])
\end{lstlisting}

\begin{lstlisting}[language=Python]
[[ 0.46838854, -0.03855411,  1.        ],
[-0.04176139,  0.46838854, -0.48478635],
[ 0.16611878, -0.07434725,  0.46838854]]
\end{lstlisting}

\begin{figure}[H]
\centering
\includegraphics[width=70mm]{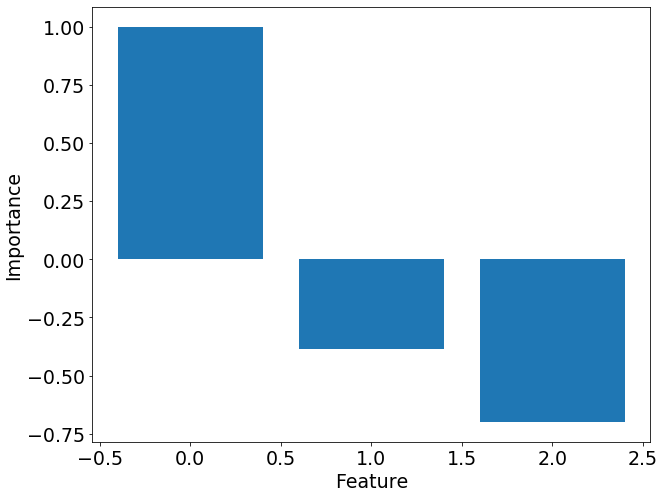}
\caption{\label{fig:mp} Features importance for Iris data set.}
\end{figure}

The function returns a list of dictionaries (one dictionary for each input data set) with keys representing hyperparameters used in the learning process, weight matrix, training error and importance of the features, which is shown in Figure~\ref{fig:mp}. Note that in Iris dataset, feature 0 is the most important feature wherese feature 1 is the least significant one in the decision-making.

\section{Scenario analysis with FCMs} \label{sec:ScenarioFCM}

Scenario analysis in an FCM framework is often implemented by either changing the baseline values of the concepts (single shot interventions) or by introducing the proposed scenario as a new factor in the defined FCM and specifying the causal impact the proposed intervention has on the target concepts (continuous interventions). The single shot interventions mimic interventions which stop when a desired change in the specific target variables are achieved. In the continuous case, the intervention becomes part of the system and continuously impacts the target variables \cite{giabbanelli2014creating}. The \textit{Intervention} module provides the respective methods for analyzing different intervention cases. The module is instantiated by passing a simulator object to the constructor.

\begin{lstlisting}[language=Python]
>>> from fcmpy import FcmSimulator, FcmIntervention
>>> inter = FcmIntervention(FcmSimulator)
\end{lstlisting}

Before specifying intervention cases and running simulations for each scenario, we need to create the baseline for the comparison (i.e., run a simulation with baseline initial conditions and take the final state vector). To do this one needs to call {\ttfamily initialize()}  method.

\begin{lstlisting}[language=Python]
>>> inter.initialize(initial_state=init_state, weight_matrix=weight_matrix, 
                        transfer='sigmoid', inference='mKosko',
                        thresh=0.001, iterations=50, l=1)
\end{lstlisting}

\begin{lstlisting}[language=Python]
The values converged in the 7 state (e <= 0.001)
\end{lstlisting}

We can inspect the results of the initial simulation run (i.e., `baseline') in the {\ttfamily $test\_results$} field as follows:

\begin{lstlisting}[language=Python]
>>> inter.test_results['baseline']
\end{lstlisting}

\begin{lstlisting}[language=Python]
        C1        C2        C3        C4        C5        C6        C7        C8
0  1.000000  1.000000  0.000000  0.000000  0.000000  0.000000  0.000000  0.000000
1  0.750260  0.731059  0.645656  0.710950  0.500000  0.500000  0.549834  0.785835
2  0.738141  0.765490  0.749475  0.799982  0.746700  0.769999  0.838315  0.921361
3  0.730236  0.784168  0.767163  0.812191  0.805531  0.829309  0.898379  0.950172
4  0.727059  0.789378  0.769467  0.812967  0.816974  0.838759  0.908173  0.954927
5  0.726125  0.790510  0.769538  0.812650  0.818986  0.839860  0.909707  0.955666
6  0.725885  0.790706  0.769451  0.812473  0.819294  0.839901  0.909940  0.955774

\end{lstlisting}

We can use the {\ttfamily $add\_intervention()$} method to specify the intervention cases. To specify a single shot intervention we must specify the name of the intervention and supply new initial states for the concept values as a dictionary (see the code snippet below).

\begin{lstlisting}[language=Python]
>>> inter.add_intervention('intervention_1', type='single_shot', 
                        initial_state = {'C1': 0.9, 'C2' : 0.4})
\end{lstlisting}

For continuous intervention cases we must specify the name of the intervention, the concepts the intervention targets and the impact the intervention has on these concepts. In some cases we might be interested in checking scenarios where the intervention fails to be delivered to its fullest. For such cases we can specify the effectiveness of a given intervention case by setting the (optional) effectiveness argument to a number in the [0,1] interval (see the code snippet below). The effectiveness will decrease the \textit{expected} causal strength of the intervention: for example, if an intervention is expected to reduce stress by 0.5 but is only 20\% effective, then its actual reduction will be 0.1.

\begin{lstlisting}[language=Python]
>>> inter.add_intervention('intervention_1', type='continuous', 
                        weights={'C1':-.3, 'C2' : .5}, effectiveness=1)
>>> inter.add_intervention('intervention_2', type='continuous', 
                        weights={'C4':-.5}, effectiveness=1)
>>> inter.add_intervention('intervention_3', type='continuous',
                        weights={'C5':-1}, effectiveness=1)
\end{lstlisting}

In the example above, we specify three intervention cases. The first intervention targets concepts (nodes) C1 and C2. It negatively impacts concept C1 (-0.3) while positively impacting the concept C2 (0.5). We consider a case where the intervention has maximum effectiveness. The other two interventions follow the same logic but impact other nodes.

After specifying the proposed interventions, we can use the {\ttfamily test\_interventions()} method to test the effect of each case. The method requires the name of the intervention to be tested. Users also have the possibility of changing the number of iterations for the simulation; its default value is the same as specified in the initialization (see the code snippet below).

\begin{lstlisting}[language=Python]
>>> inter.test_intervention('intervention_1', iterations=10)
>>> inter.test_intervention('intervention_2')
>>> inter.test_intervention('intervention_3')
\end{lstlisting}

\begin{lstlisting}[language=Python]
The values converged in the 7 state (e <= 0.001)
The values converged in the 6 state (e <= 0.001)
The values converged in the 7 state (e <= 0.001)
The values converged in the 7 state (e <= 0.001)
\end{lstlisting}

\begin{figure}[H]
\centering
\includegraphics[width = \textwidth]{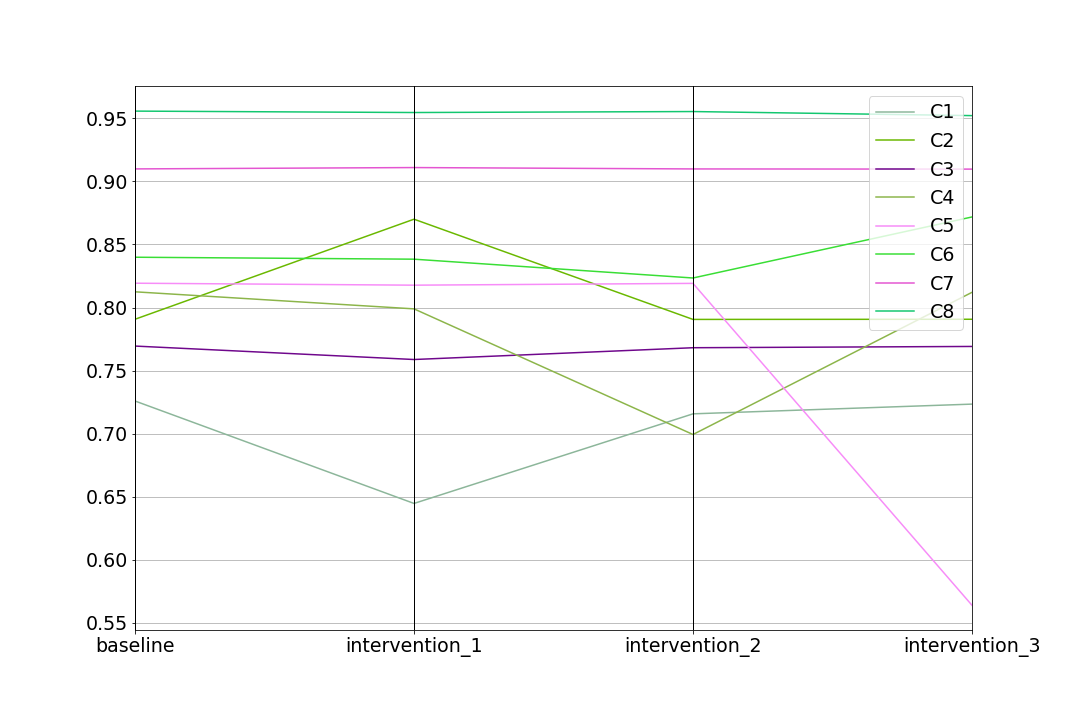}
\caption{\label{fig:inter} Baseline and intervention scenarios.}
\end{figure}

The equilibrium states of the interventions can be inspected in the {\ttfamily equilibriums} field (see Figure \ref{fig:inter} and the code snippet below).

\begin{lstlisting}[language=Python]
>>> inter.equilibriums
\end{lstlisting}

\begin{lstlisting}[language=Python]
    baseline    intervention_1  intervention_2  intervention_3
C1  0.725885        0.644651        0.715704        0.723417
C2  0.790706        0.870060        0.790580        0.790708
C3  0.769451        0.758786        0.768132        0.769141
C4  0.812473        0.798947        0.699316        0.812073
C5  0.819294        0.817735        0.819160        0.563879
C6  0.839901        0.838350        0.823430        0.871834
C7  0.909940        0.911004        0.909917        0.909778
C8  0.955774        0.954652        0.955427        0.952199
\end{lstlisting}

Lastly, one can inspect the differences between the interventions in relative terms (i.e., $\%$ increase or decrease) compared to the baseline (see the code snippet below).

\begin{lstlisting}[language=Python]
>>> inter.comparison_table
\end{lstlisting}

\begin{lstlisting}[language=Python]
    baseline    intervention_1  intervention_2  intervention_3
C1       0.0      -11.191083       -1.402511       -0.339981
C2       0.0       10.035821       -0.015968        0.000202
C3       0.0       -1.385998       -0.171325       -0.040271
C4       0.0       -1.664794      -13.927524       -0.049314
C5       0.0       -0.190233       -0.016379      -31.175022
C6       0.0       -0.184640       -1.960979        3.802010
C7       0.0        0.116873       -0.002543       -0.017806
C8       0.0       -0.117365       -0.036331       -0.374038
\end{lstlisting}

%% -- Summary/conclusions/discussion -------------------------------------------

\section{Conclusions and future research} \label{sec:summary}
The \textbf{FCMpy} package provides a complete set of functions necessary to conduct projects involving FCMs. We created a tool that is open-source, easy to use, and provides the necessary functionality. The design and implementation of the tool results from a collaboration with multiple experts from the field of FCMs. We believe that this tool will facilitate research and encourage new students and scientists to involve FCMs in their projects. 

We included both well-known algorithms as well as recently developed ones. We are planning to constantly update our library and welcome all scientific community contributions.

\section*{Acknowledgments}
Developed Python code: Mkhitaryan and Wozniak. Contributed to the development of
the classification algorithms: Nápoles. Designed the project: Mkhitaryan, Wozniak, Giabbanelli, Crutzen. Supervised the project: Giabbanelli, Crutzen. Wrote the first draft: Mkhitaryan, Wozniak, Giabbanelli. Edited and approved the manuscript: all authors.

%% -- Bibliography -------------------------------------------------------------
%% - References need to be provided in a .bib BibTeX database.
%% - All references should be made with \cite, \citet, \cite, \citealp etc.
%%   (and never hard-coded). See the FAQ for details.
%% - JSS-specific markup (\textit, \textit, \code) should be used in the .bib.
%% - Titles in the .bib should be in title case.
%% - DOIs should be included where available.
\bibliographystyle{unsrt}  
\bibliography{article_archivx} 
\end{document}